\font\gross=cmr10 scaled \magstep1
\font\mengen=bbm10
\font\ninebf=cmr7
\def\doppel{\magnification=\magstep1}
\def\MZ{\hbox{\mengen Z}}
\def\MR{\hbox{\mengen R}}

\def\MC{\hbox{\mengen C}}
\def\MQ{\hbox{\mengen Q}}
\newcount\figcount \figcount=0
\def\fig#1#2{\global\advance\figcount by1\midinsert\vskip #2
\centerline{Fig.\ \the\figcount : #1}\endinsert}
\def\psfig#1#2{\global\advance\figcount
by 1 \midinsert\vbox{\centerline{\epsfbox{#2}}
\centerline{Fig.\ \the\figcount : #1}}\endinsert}
\def\foryoureyesonly#1{}
%
\def\paper#1#2#3#4#5#6#7#8#9{\item{\bf [#1]}#2: ``#3'', #4 {\bf
#5} (#6) p.\ #7 #8 \foryoureyesonly{#9}\par\goodbreak}
%
%

%
%
\def\eprint#1#2#3#4#5#6{\item{\bf [#1]}#2: ``#3'', {{\tt #4}} #5
\foryoureyesonly{#6}\par}
%
%

%
%
\newcount\chapterno
\newcount\subchapterno
\newcount\glno
\edef\rememberref#1{\relax}  
\def\rememberchapter#1{\relax}  
\def\remembersubchapter#1{\relax}  
\def\neueseitevorchapter{\vfill\supereject\ifodd\pageno\relax\else \noindent$
$\vfill\eject\fi} 
\def\kapitelnummer{\the\chapterno}
\def\chapter#1{\neueseitevorchapter\global\advance\chapterno by
  1\expandafter\rememberchapter{\the\chapterno . #1}\glno =0
  \subchapterno =0\titel{\the\chapterno . #1}} 
\def\subchapter#1{\global\advance\subchapterno by
    1\remembersubchapter{\the\chapterno .\the\subchapterno
    . #1}\subtitel{\the\chapterno .\the\subchapterno . #1}}
\def\gln#1{\global\advance\glno by 1\xdef#1{(\the\chapterno
.\the\glno)}\eqno {(\the\chapterno .\the\glno)\ifdraft\hbox to 0cm{\tt\string #1\hss}\else\relax\fi}}
\newif\ifdraft
\draftfalse
\def\draft{\drafttrue\def\vielluft{\bigskip}\def\neueseitevorchapter{\par}\def\foryoureyesonly##1{##1}\def\shlabel##1{\hbox
  to 0cm{\tt \expandafter\string ##1\hss}}}
\def\openbib{\def\aux{1}\openout\aux=\jobname.aux \immediate\write16{Putting
    references in \jobname.aux}
    \def\rememberref##1{\write\aux{cite(##1) on page \folio}}
    \def\rememberchapter##1{{\edef\x{\noexpand\write\aux{chapter[##1] on page
    \noexpand\folio}}\x}}
    \def\remembersubchapter##1{\write\aux{subchapter[##1] on page \folio}} }
\def\closebib{\closeout\aux \def\rememberref##1{\relax}}
\def\cite#1{\raise 0.5ex\hbox{\ninebf[#1]}\rememberref{#1}}
\def\titel#1#2{{\removelastskip\bigskip\goodbreak\noindent\mark{#1}\gross
#1\medskip \nobreak\noindent #2\unskip}}
\input epsf
\doppel
\openbib
\def\neueseitevorchapter{\par}
\parskip=1ex

\centerline{\gross Lessons from the LQG String}
\centerline{Robert C. Helling}
\centerline{International University Bremen}
\centerline{\tt helling@atdotde.de}
\bigskip
{\bf Abstract:}

We give a non-technical description of the differences of quantisation
of the bosonic string between the usual Fock-space approach and the
treatment inspired by methods of loop quantum gravity termed the LCQ
string. We point out the role of covariant states with continuous
representations of the Weyl operators versus invariant states leading
to discontinuous polymer representations. In the example of the
harmonic oscillator we compare the optical absorption spectrum for the
two quantisations and find that the question of distinguishability
depends on the order in which limits are taken: For a fixed UV cut-off
restricting the Hilbert space to a finite dimensional subspace the
spectra can be made arbitrarily similar by an appropriate choice of
state. However, if the states are chosen first, they differ at high
frequencies.

\chapter{Introduction}
Recently, there have been a number of attempts to bridge the gap of
understanding between the string and the loop quantum gravity
community. After a ``Strings meet Loops'' conference at the AEI in
Golm, there had been the paper by Thiemann\cite{T1}
discussing the quantisation of the bosonic string with loop inspired
techniques with a number of surprising results. As a follow-up we
published \cite{HP} where we contrasted the construction of \cite{T1}
with the usual Fock space construction in a common notational
framework focussing on the issue of the conformal anomaly or central
charge.

In an attempt to make precise statements, \cite{HP} is to some extend
quite technical. As a result, there are a number of misconceptions
about that work. In many discussions I learned that we should have
been more careful to point out the central points of \cite{HP} and
contrasting them to mere technical details. Furthermore, my own
understanding of several of the issues involved has grown in the last
two years. Therefore I feel it might be profitable to publish this
note to explain in a much less technical manner the differences
between the two approaches to the bosonic string and possible
generalisations to other theories with gauge (e.g. diffeomorphism)
symmetries such as gravity. On the other hand, this means that the
reader interested in details of arguments that have been left out for
readability here should consult \cite{HP}.

Even before I start, I would like to point out that all the issues
discussed in \cite{HP} and here only apply at the ``kinematical''
level (i.e. the treatment of spacial diffeomorphisms in loop
quantum gravity). As the world-sheet theory of the string splits into
independent left- and right-moving (holomorphic and anti-holomorphic)
sectors it becomes effectively one dimensional and this kinematical treatment
is all that is required. In higher dimensional theories, after one has
solved these kinematical issues, one can attack the much thornier
problems of the dynamics that appear around the treatment of the
Hamiltonian constraint. I have nothing to say about these and the
interested reader should consult \cite{NPZ} and \cite{NP} where these
are discussed in detail from a stringy perspective and \cite{T2} for
the ``inside'' perspective. In this sense, our discussion is orthogonal to
what is discussed in those papers.

The structure of this note is as follows: In the following section, we
review the two approaches to the quantisation of the bosonic
string. The main lesson from this is the distinction between invariance
and covariance of the state to build the Hilbert space representation
upon. Then, in order to understand the physical consequences of the
discontinuity of the polymer representation we shift attention to the
harmonic oscillator and discuss the optical absorption spectrum in the
two treatments. There is a final section before conclusions where the
philosophical details for the possibility of a distinction are
detailed. Two appendices contain the technical details of the
calculation of the absorption spectra for a single and several
oscillators. 

This note grew out from discussions of \cite{HP} with a number of
people to which my thanks are due. These include Aaron Bergmann, Klaus
Fredenhagen, Bernard de Wit, Jacques Distler, Josh Gray, Michael
Green, Lubos Motl, Hermann Nicolai, Hendryk Pfeiffer, Kasper Peeters,
Giuseppe Policastro, Thomas Thiemann, and Maria Zamaklar.

\chapter{The bosonic string in a nutshell}
In this chapter, I will summarise the core of the construction of
\cite{T1} and \cite{HP} for reference in the later discussion. 

To understand the issues in which the two approaches to the
quantisation of the bosonic string differ one has to distinguish
between the abstract algebra $\cal A$ of observables and its
representation as operators acting on a Hilbert space. In the first
step, both approaches are parallel and only in the second, the choice
of the Hilbert space, there is a difference with potentially physical
consequences.

The algebra $\cal A$ is generated by elements denoted $W(f)$ where
$f\colon S^1\to\MR$ is a real function on the spacial (actually:
light-like) part of the world-sheet of the string which for the closed
string we are discussing is a circle. Although one should think of $f$
as a test-function used to smear the current $\partial X$, we will not
worry about smoothness properties of these functions as in the end the
algebra will be completed to a C*-algebra anyway and thus it is enough
to work on a dense subset. In \cite{T1}, Thiemann uses piecewise
constant functions as these can be thought of analogues of holonomies
along paths (or spin networks in the higher dimensional case) which
are central in the LQG framework while in the usual textbook treatment
of the string, $f$ is expanded in its Fourier modes (usually denoted
as $\alpha_n$). Here, we will use general functions $f$ to have a
unified language.

The product of two such operators $W(f)$ and $W(g)$ is given by
$W(f+g)$ up to a phase proportional to
$\sigma(f,g)=\int_{S^1}fdg$. Expanded in Fourier modes, this phase is
the representation of the canonical commutation relations
$[\alpha_m,\alpha_n] = m\delta_{m+n}$. Note well that the expression
for the phase is given in coordinate free form. Thus it is independent
of any choice of coordinates. We can reformulate this as an active
rather than passive transformation as $\sigma(f,g)=\sigma(f\circ S,
g\circ S)$ in terms of diffeomorphisms $S:S^1\to S^1$ which map the
circle onto itself. This is the diffeomorphism invariance that we
would like to preserve in the quantum theory as well.

In any case, these diffeomorphisms are symmetries (automorphisms more
precisely) of the algebra $\cal A$. The crucial question is if they
are also symmetries of the Hilbert space on which $\cal A$ is
represented. Specifically, for each such $S$ we need a unitary
operator $U(S)$ on the Hilbert space such that for the operators
$\varrho(W(F))$ (here $\varrho:{\cal A}\to End({\cal H})$ is the
representation) we have
$$\varrho(W(f\circ S))=U(S)^{-1}\varrho(W(f))U(S).\eqno(1)$$ 
The existence of these $U(S)$ is really what makes the Hilbert space
covariant. 

Of course there are many diffeomorphisms and they form a group under
concatenations. For two such diffeomorphisms $S$ and $T$, (1)
guaranties that $U(S)U(T)=U(S\circ T)\times\exp(i\phi(S,T))$ for some phase
$\phi(S,T)$. Obviously the relation (1) only defines $U(S)$ up to a
phase $\psi(S)$ and making a different choice for the $\psi$'s changes
$\phi$ as well. In the following, one has to proceed to the physical
Hilbert space of invariants of all $U(S)$. This should not be $\{0\}$
and therefore one needs to find an assignment of $\psi$'s that make
all the $\phi$'s vanish. Unfortunately, this is not always possible
and the obstruction is the central charge or conformal anomaly.

For example, in the usual Fock space construction of the $U(S)$ (which
is a bit lengthy and described in detail in \cite{HP}) for a single
target space coordinate $X$ the central charge does not
vanish. However, one can take $D$ copies of the algebra $\cal A$ and
represent them on the direct sum of Hilbert spaces. The central charge
is additive under this addition of theories and it turns out that by
taking exactly $D=26$ copies of the theory and adding to it a similar
but fermionic theory of the $bc$-ghost system one can make the total
phase $\phi$ vanish. This is the way the critical dimension appears in
this setup.

To understand the statement that in the LQG quantization there is no
anomaly and thus no central charge one has to understand the
connection between states of the algebra $\cal A$ and its
representations. For Hilbert space representations of C*-algebras are
constructed in a similar way (called GNS construction) as highest
weight representations of Lie algebras: One starts with a state
(called the vacuum but as we are not discussing dynamics there is no
notion of energy and thus it should not be thought of as ``lowest
energy state'' but just as a reference state) and acts on it with the
elements of $\cal A$. This state is characterised by specifying the
expectation values in this state of all the observables in the algebra
$\cal A$. Formally, it is given by a linear function $\omega:{\cal
A}\to \MC$. It has to be positive and normalised as $\omega(W(0))=1$
as $W(f)$ for the function $f$ which is 0 everywhere is the unit in
$\cal A$. The Hilbert space built on this $\omega$ contains besides
the vacuum $|0\rangle$ all the vectors $|A\rangle$ created by acting
with elements $A\in{\cal A}$ on that vacuum. The role of $\omega$ is
to supply this Hilbert space with a scalar product as $\langle
A|B\rangle = \omega(A^*B)$ where the adjoint of an $W(f)$ is given by
$W(f)^*=W(-f)$.

As any element in $\cal A$ can be given as a linear combination of
$W(f)$'s it suffices to give $\omega$ just on those observables. 
Every Hilbert space representation can be obtained in this way. However,
we have not yet constructed the unitary implementers of the symmetries
$U(S)$. If the state $\omega$ happens to be invariant under the
symmetries, that is $\omega(W(f))=\omega(W(f\circ S))$ for all $f$,
there is an easy choice: Just define $U(S)$ as the operator that maps
$|W(f)\rangle$ to $|W(f\circ S)\rangle$. The invariance of $\omega$
implies that this map is unitary. Furthermore, with this choice of
$U(S)$ the phase $\phi(S,T)$ which would lead to the central charge
vanishes automatically. But we should reiterate that this simple choice is
only available if $\omega$ is invariant as otherwise these operators
fail to be unitary.

In fact, a map $\omega$ such that $\omega(W(f))$ which is invariant
under moving the points of the $S^1$ around cannot locally depend on
$f$: The choice according to LQG (the ``polymer state'') accordingly
is to set $\omega(W(f))$ to 1 if $f$ is the function $f=0$ which
vanishes everywhere and to 0 otherwise. The price for this invariance
however is a high one: This $\omega(W(f))$ is obviously not continuous
under variations of $f$! We will explore the physical consequences of
this discontinuity in the next chapter.

In contrast, in the usual Fock space quantisation of the string, the
vacuum is in general {\sl not} invariant under
diffeomorphisms. Rather, it is only invariant under those
diffeomorphisms which do not mix positive and negative frequency
modes. Those diffeomorphisms (acting as Boguliubov operators) which
map particles to anti-particles and vice versa change the Fock
vacuum. Therefore for those $S$ one has to define $U(S)$ by different
means which are detailed in \cite{HP}. The price to pay however is the
central charge which selects the critical dimension as explained above.

\chapter{Invariance vs. Covariance}
It seems, both quantisations have their disadvantages: The Fock space
quantisation is anomalous while the LCQ inspired quantisation has a
discontinuity. 

Indeed, the anomaly of the Fock quantisation fatal: If present, the
physical Hilbert space of invariants is empty. However in the critical
dimension it cancels between the space-time fields and the
ghosts. Thus in this case, the bosonic string provides a non-trivial,
continuous and covariant (in the sense that the $U(S)$ exist for all
diffeomorphisms) representation. This construction (which is in fact
the Gupta-Bleuler construction in disguise) yields a realisation of
gauge symmetries which is potentially anomalous and thus only works
for special classical theories where the anomalies happen to
vanish. The other price to pay is that (in some sense half of) the
unitary implementers $U(S)$ act in a non-trivial way on the vacuum and
are thus spontaneously broken in this state.

This is in contrast to the LQG quantisation where the vacuum is
invariant by construction and as we have explained there is never an
anomaly. Here the price is the discontinuity of the representation of
the algebra which results in a quite singular scalar product: In fact,
two states $|W(f)\rangle$ and $|W(g)\rangle$ only have a non-vanishing
scalar product (and thus overlap) if $f=g$. This induces the discrete
topology on the space of functions $f$. Not only diffeomorphisms but
all maps $\tilde S$ which point-wise map $S^1$ to itself are
symmetries. Thus any topological notion of ``nearness'' is lost and
only ``equality'' is preserved.

In higher dimensions, there is a theorem\cite{LOST} which shows that
the requirement of invariance of $\omega$ is so strong that it singles
out the polymer state uniquely. It seems plausible that a similar
uniqueness theorem holds in the case of the string as well. This
theorem is usually cited as justification for the inevitability of the
polymer state in LQG.

The existence of the Fock space is not a counter example to this
possible generalisation as we argued above it is not
invariant. However, as explained, the physical requirement is only the
existence of the unitary implementers $U(S)$ and those happen to be
particularly simple for Hilbert spaces built on invariant states. The
Fock space is an example where the $U(S)$ are realised in a different,
non-trivial way. What this example really shows that other unitary
implementers than those from invariant states can have better
properties as they allow for a continuous representation of the
algebra. Thus the restriction of attention to Hilbert spaces built on
invariant states is at least not the most general one and might in
fact be too narrow. Unfortunately, constructing $U(S)$ in general is a
very hard problem and there is no solution in sight for the more
relevant case of higher dimensional diffeomorphisms needed for
realistic theories of quantum gravity. What the example of the bosonic
string suggests is that consequences of the discontinuity of the
polymer state might be in fact due to a poor choice of Hilbert space
representation which is not mathematically forced but just a choice of
convenience. It might be that there are better choices where some of
the diffeomorphisms are spontaneously broken.

One should compare this to the situation in classical general
relativity: GR is obviously diffeomorphism invariant. However
any classical state given by Cauchy data or a four-metric which
satisfies Einstein's equations breaks an infinite number of
diffeomorphisms and just leaves a finite number of isometries
unbroken. For the Minkowski solution there are just four translations
three rotations and three boosts.

Of course, it is still possible to use different coordinate systems in
Minkowski space such as spherical coordinates and thus the theory is
still diffeomorphism invariant. The important point is however that
the solution specified by the metric tensor is not invariant but the
metric has to be transformed. One has to take the pull-back!
Similarly, in general, the state represented by $\omega$ has to be
transformed when a diffeomorphism is applied. And this is achieved by
the non-trivial $U(S)$.

In fact, there is one choice of tensor which is invariant under all
diffeomorphisms which is $g_{\mu\nu}=0$. This tensor is of course not
suitable as a metric as it fails to be invertible. Requiring an
invariant state in classical relativity would force one to choose this
pathological metric. 

So, there might be other choices of covariant Hilbert spaces for
quantum gravity than those built upon the polymer state just as the
bosonic string has a continuous representation which is not the
polymer representation. But there remains the question if one should
regard the polymer state as a valid (although maybe unfortunately
chosen) physical state or whether one should abandon it because of the
discontinuity. In the end, \cite{T1} proposes it as a quantisation of
the string in others than the critical dimension.

Thus, the question is if one should regard weak continuity as a physical
requirement of a ``good'' quantisation procedure or if we can do
without. Experience has shown that in the past, successful
quantisations of physical theories there was a continuous
representation available and one did not need to resort to these more
singular choices. But before a fully formulated theory of quantum
gravity is known there is the possibility that for such a theory there
are no other choices available.

\chapter{Weak discontinuity: Harmonic Oscillator}
At the current state, one can still try to understand the physical
consequences of the discontinuity. In \cite{HP} we approached this
question by considering an even simpler but physically well known
system where we explored the properties of a discontinuous
quantisation along the lines of the polymer state: The harmonic
oscillator.

At the kinematical level, we are talking about quantum mechanics of a
single degree of freedom. There, the algebra of Weyl operators looks
very similar to the algebra for the bosonic string, the only
difference being that one has to replace $W(f)$ by $W(z)$ where $z$ is
a complex number representing a point in phase space. The
multiplication law is given by $W(z_1)W(z_2)=W(z_1+z_2)\exp(i \Im(\bar
z_1 z_2))$. The usual Fock space vacuum is represented by
$\omega_F(W(z))=\exp(-|z|/4)$ while again the polymer state is 1 for
$z=0$ and vanishes elsewhere.

The first physical observation about this polymer state is that
formally it can be obtained as a thermal state: The KMS-state one
obtains by coupling the oscillator to a heat-bath is labelled by the
inverse temperature $\beta$ (which determines the periodicity in
imaginary time). Taking the limit $\beta\to 0$ yields the polymer
state. Thus this state can be thought of as a thermal state of
infinite temperature. It is not surprising that this state has unusual
spectral properties. 

The crucial difference to the string is that there are no gauge
symmetries one wishes to mod out. However, the time evolution
characterising the harmonic oscillator (equivalent to giving the
Hamiltonian if that existed) has a very similar form as it
is given by $\alpha_t(W(z))=W(e^{it}z)$. This time, both states are invariant
under this transformation and thus the unitary time evolution operator
on the respective Hilbert spaces $U(t)$ is readily available.

In the Fock representation, it is continuous in $t$ and one can define
an unbounded operator $H={d\over dt}U(t)|_{t=0}$ which has the well
known evenly spaced spectrum that one can for example observe in Raman
molecular vibration spectra. Furthermore, there are stationary states
in the Hilbert space which are labelled by an integer, the excitation
number.

In the polymer representation, due to the discontinuity the derivative
does not exist thus there is no Hamiltonian and it does not make sense
to ask for its spectrum. This has a parallel in the polymer string
where one cannot take derivatives which would yield operators for the
field and the field momentum or creation and annihilation
operators. Only the integrated finite transformations exist.

In \cite{HP}, we defined a family of operators $H_\epsilon$ which are
finite difference versions of the time derivative of the time
evolution $U(t)$. In Fock space, where the derivative exists,
$H_\epsilon$ becomes the Hamiltonian when $\epsilon\to0$. In the
polymer Hilbert space we showed that generically the $H_\epsilon$ have
vanishing expectation value and a variance which for a fixed state
scales as $1/\epsilon^2$. In no Fock state, the $H_\epsilon$ have
these properties. This we took as a clear indication of different
physics of the two quantisations.

This conclusion was criticised on the ground that it is physically
impossible to probe arbitrarily short intervals of time (for example
the Planck time being a natural cut-off) and thus taking the limit
$\epsilon\to0$ is not physical. For any finite $\epsilon$ the variance
is finite and thus one does not get a sharp distinction.

\indent\cite{HP} only discusses properties of a decoupled harmonic
oscillator. In this note, we take the analysis further by coupling the
oscillator to its environment to ``observe'' its properties.

The polymer Hilbert space is much bigger (it is not separable) than
Fock space as all vectors $|W(z)\rangle$ are orthogonal to each other
and have no overlap. This implies that the subspace orthogonal to a
given vector is in some sense much larger: Take an eigenvector of some
Hamilton operator in a separable Hilbert space. Then act on it with a
generic operator describing an interaction. The resulting vector
typically has non-zero overlap with all other eigenvectors of the
Hamiltonian unless there are selection rules forbidding a certain
transition.

This is different in the polymer Hilbert space: There, the overlap can
only be with a subspace of countable dimension in the whole space of
uncountable dimension. Thus, generic transitions are always forbidden
for a given interaction.

Even if one focusses on those special pairs of states for which
interactions are possible for a given perturbation, the resulting
picture differs. In the appendix, we compute the absorption spectrum
for the polymer harmonic oscillator interacting with electromagnetic
radiation of frequency $\Omega$. We find the rate of absorption to be
given by
$$A_P = {1\over8\sin^2\left(\pi (m'-m-\Omega)\over N\right)}.$$
$N$ is an arbitrary integer assumed to be large but finite. It is a
property of the pair of states involved in the transition.  $m$ and
$m'$ are ``generalised frequencies'' of the two states involved with
the condition that both have to divide $N$. This should be compared to
the rate of absorption computed in the ordinary Fock Hilbert space:
$$A_F= {1\over 2 (\omega_{m'}-\omega_m+\Omega)^2}.$$ As detailed in
the appendix, the question of whether the two expressions can be
distinguished depends on the order of limits to be taken (there is a
trivial factor of $(2\pi/N)^2$ which could be absorbed in in the
``path integral'' over the time of interaction): For
$|\Omega|<\Omega_{max}$ for fixed $\Omega_{max}$, the two expressions
can be made arbitrarily similar by taking a sufficiently large
$N$. However, for any fixed $N$ (that is for fixed pair of states),
the two expressions deviate for large enough $\Omega$. 

This behaviour is typical for comparisons between Fock- and
polymer-physics: For a spe\-ci\-fied set of measurements, it is possible
to find states in the polymer prescription such that for these
measurements the two approaches agree. However, by performing
measurements outside the predetermined set, it is possible to
differentiate the two approaches. Whether this means that the to
approaches are physically equivalent or not is thus a philosophical
question as it depends on the exact rules of the comparison: Do we
have to prepare the system before we decide which measurement to make
to make or is it allowed to prepare the system afterwards in a state
in the polymer Hilbert space that mimics the Fock state for that
specific measurement?

\chapter{Can we observationally distinguish Fock- and the polymer Hilbert-space}
Remember $N$ was chosen as an arbitrary integer above. So what is its
physical status? There are two basic strategies to reconcile the two
results. Either $N$ is to be understood as a fixed physical constant
{\sl a priori} or it is chosen {\sl a posteriori} according to
circumstances. Both have parallels in the literature.

The first alludes to ideas of ``space-time foam'' and posits that at
the Planck scale \hbox{(space-)} time has a fundamental granularity and it is
only possible to measure time in integer multiples of Planck
time. Then $N$ is the oscillatory period of the oscillator in Planck
units. This would imply that it is physically impossible to access
frequencies larger than $N$ and thus the periodicity of the absorption
would be invisible in principle. Furthermore, all our measurements of
oscillators in the past would have been in the $m,m',\Omega\ll N$
regime and thus we could not have observed a deviation from $A_F$. In
this case however the spectra of oscillators with slightly
different frequencies would look very different depending on the
selection rule that restricts absorption to possible values for $m-m'$
in terms of divisors of $N$. This is not a very attractive scenario.

Furthermore, as time in that world would be quantised in units of
$2\pi/N$, the effective Hilbert space would separate into subspaces
each of dimension $N$ rendering it effectively finite dimensional. In
this restriction, and time being discrete, anything becomes
continuous (as for the discrete space-time there is only the discrete
topology). Thus by restricting to a finite dimensional subspace it is
not surprising that there are no differences between the two
constructions which differ in their continuity properties. What this
approach does is to build a hard UV regulator into the theory. This of
course is a possible although not very popular approach which could as
well be used in theories of quantum gravity immediately solving
problems associated to non-renormalisability. 

Even if this is not a problem in the context of the oscillator, it
appears to be very difficult to implement a UV cut-off in a
diffeomorphism invariant way. So this possibility might not be
available in the case of the string or gravity.

The other possibility is to choose $N$ on an ad hoc basis depending on
the measurements one is going to perform. For example you intend to
match observations of absorption lines done with a certain
bandwidth. Then you can choose $N$ sufficiently large to prepare the
oscillator in a state $|\psi_m\rangle$ for this $N$ such that
deviations from the absorption spectrum of a Fock oscillator are
invisible at this bandwidth.

If $N$ is not tied to a constant of nature (like the Planck time), one
could also consider transitions between states of different $N$, say
$N_1$ and $N_2$. In this case however, the two scalar products
involved in the computation of $\cal M$ would only be simultaneously
non-vanishing if $t'$ is an integer multiple both of $2\pi/N_1$ and
$2\pi/N_2$. Thus this computation is equivalent to a transition where
both initial and final state have a common $N_{eff}
=\gcd(N_1,N_2)$. As the periodicity in the absorption rate becomes
visible for $\Omega \sim N_{eff}$, considering transitions with different
$N$ would lower the effective $N$ and thus increase the deviation from
the Fock space result.

In this second approach, there is no fundamental granularity of time and $N$
is just a property of the quantum state the oscillator happens to be
in. On the other hand, in this approach, as there is no fundamental limit on
possible observations, one could, after the state is prepared, decide
to measure outside the prespecified bandwidth and reveal the
difference of that oscillator from the one using Fock space.

Again, one could object that any real world oscillator has its built
in bandwidth outside which it ceases to behave like an harmonic
oscillator. For example, once too much energy has been put into a
vibrational mode of a molecule, the oscillatory degree of freedom is
no longer decoupled from the remaining degrees of freedom (including
eventually fundamental particle degrees of freedom inside the nuclei)
and ceases to be harmonic. However, only slightly extending this
argument against analysing idealised subsystems one should never
discuss systems like the harmonic oscillator and only attempt to
quantise the universe as a whole interacting system as anything else
is an unphysical idealisation.

In the case of two coupled oscillators (see appendix) one sees
that in order to make the eventually decreasing absorption
unobservable one would have also to forbid arbitrarily long
measurements. Thus in addition to a UV regulator one as well has to
impose an IR regulator and choose the state accordingly. Therefore an
analogous reasoning applies: What has to be chosen first, the state or
the regulator on the observations?

If one decides to follow the first assumption above and assume a
fundamental (Planck) frequency such that all oscillators would have to
have frequencies $1/N$ times that fundamental frequency for some $N$
this would have to hold for the eigenfrequencies as well and
the irrationality assumption of the eigenfrequencies would be impossible
to fulfil. This however would impose quantisation conditions on the
allowed range of possible couplings $\lambda$ on which the
eigenfrequencies depend non-polynomially as well which appears very
unnatural. 

\chapter{Conclusions}
This note contrasted two approaches to quantisation, especially in
their application to the world-sheet theory of the bosonic string. The
polymer state is based on an invariant state while the Fock state is
only covariant and thus potentially anomalous and in fact the anomaly
vanishes only in the critical dimension. This example shows that
requiring invariance of the state in the quantisation of a gauge
system is too strict and one should in fact only ask for covariance
(existence of unitary implementers of the symmetry).

As a result of this too strong invariance requirement the polymer
state ceases to be continuous and thus leads to a very singular
Hilbert space. The second half of this note investigated physical
consequences of this discontinuity. Specifically, in the example of
the harmonic oscillator, optical absorption spectra were computed in
perturbation theory. It was shown that for a given finite bandwidth
spectrum one can construct states in the polymer Hilbert space which
resembled the Fock result to an arbitrary finite precision. Once
prepared however, the absorption differs significantly at high
frequencies outside the prespecified range. Thus for any state in the
polymer Hilbert space can be observationally distinguished from the
usual Fock states assuming the experimenter has access to sufficiently
high frequency radiation.

\chapter{Appendix A: Absorption spectrum}
Here we compute the rate of absorption of photons from a field of
coherent radiation as a function of the frequency of that
radiation. We do this computation in first order perturbation theory
as already at this order we have good agreement with experimentally
observed absorption spectra. For the same reason the radiation field
is treated classically as QED effects should not be important for this
simple process.

The most obvious way to observe the spectrum of a physical system is
to measure an absorption spectrum. We can compute this in time
dependant perturbation theory. All we need is the time evolution, the
Hamiltonian is not required.

Let's do the textbook calculation first. The idea is to couple the
oscillator to a radiation field. More specifically, we perturb with a
``photon operator'' $M(t)=\exp(i\Omega t) |m'\rangle\langle m|$. We
start out with state $|m\rangle$, this evolves for a time $t'$
according to the free Hamiltonian. Then $M(t')$ acts and further on, the
state evolves with the free evolution until time $t$. 
\psfig{A photon induces a transition between states $m$ and
$m'$}{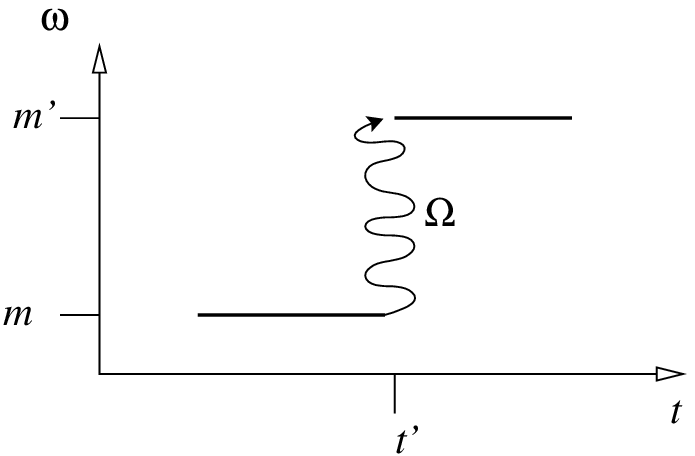}
According the
``sum over paths'' philosophy, we have to sum (integrate) all
possible $t'$. In the end, we compute the overlap with the state
$|m'\rangle$, the absolute square gives the probability of this
absorption of a photon to happen. In a formula
$$\eqalign{{\cal M} &= \langle m'|U(-t)\int_0^t dt'\, U(t-t')\exp(i\Omega
t')|m'\rangle \langle m|U(t')|m\rangle\cr
&=\int_0^t dt'\, e^{i t'(\omega_m-\omega_{m'})}\exp(i\Omega t')\cr
&= {e^{i(\omega_{m'}-\omega_m+\Omega)t}-1\over
\omega_{m'}-\omega_m+\Omega}\cr}.$$
Thus the absorption rate is given by 
$$|{\cal M}|^2 = {\sin^2((\omega_{m'}-\omega_m+\Omega) t/2)\over (\omega_{m'}-\omega_m+\Omega)^2}.$$ Therefore, with an additional time average (alternatively this can be thought of as an average over the initial phase of the radiation field) we find the absorption rate behave like
$$A_F=\langle|{\cal M}|^2\rangle = {1\over 2
(\omega_{m'}-\omega_m+\Omega)^2}.\eqno{(2)}$$
Now, we want to perform an analogous calculation for the polymer
representation. The first question is to find states similar to the
stationary states $|m\rangle$ in the usual treatment. Recall that the
states $|z\rangle$ are orthonormal ($\langle z|z'\rangle =1$ if
$z=z'$ and $0$ otherwise) for all $z\in\MC$ and the
general vector is an at most countable linear combination such that the
sum of the absolute value squares of the coefficients is
finite. The space of finite linear combinations is dense and we will
only consider those. 

For obvious reasons besides $|0\rangle$ there is no state which is an
eigenstate of $U(t)$ for all $t$ (remember the time evolution acts as
$U(t)|z\rangle = |e^{it}z\rangle$). However, it is easy to find states
which are eigenstates at $N$ instances of time during a period
$t\in[0,2\pi)$. Namely, we want to consider the normalised states
$$|\psi_m\rangle = {1\over\sqrt N}\sum_{k=1}^N e^{2\pi ikm/N} |e^{2
\pi i k/N}\rangle.$$
Here $N$ is an integer which is understood to be large but finite and
$m\in\{1,\ldots,N\}$. These states are periodic in time with period
$2\pi/\gcd(m,N)$. Thus, the angular frequency is $\gcd(m,N)$. We can
restrict ourselves to such $m$ which divide $N$ which then have
(angular) frequency $m$ in analogy to the $|m\rangle$ in Fock space.

Note that in contrast to the Fock space, the $|\psi_m\rangle$ are not
dense and their span (which is closed!) only forms a tiny part of the
polymer Hilbert space. Thus a general state cannot be decomposed into
a sum (possibly infinite) of $\psi_m\rangle$'s. Nevertheless, we
will consider these states as the closest analogues of the oscillator
eigenstates in Fock space.
 
In particular, we will concede that the radiation field couples two of
these states via the operator
$$M(t) = e^{i\Omega t} |\psi_{m'}\rangle\langle \psi_m|= e^{i\Omega t}\sum_{l=1}^N
e^{{2\pi i\over N}(m'-m)l}|e^{2\pi il\over N}\rangle\langle e^{2\pi i
l\over N}|.$$
Again, at least formally, we compute
$${\cal M} = \langle \psi_{m'}|U(-t)\int_0^t dt'\,
U(t-t')M(t)U(t')|\psi_m\rangle.$$

The integrand of the $t'$ integration vanishes almost everywhere and
is finite at discrete values in between. Thus, in a Lebesgue sense, the
integral vanishes. However, we should be careful that this integral
was a ``sum over all possible times of interaction''. Thus we silently
replace it by a sum over those $t'\in[0,t]$ where the integrand does
not vanish. For simplicity of these now discrete expressions, we
further assume that $t=2\pi P$ for some integer $P$, the number of
periods of the oscillator we shine the light on it. After some
algebra, we find the rate of absorption to be
$$|{\cal M}_{p}|^2 = {\sin^2(\pi p \Omega)\over \sin^2(\pi (m'-m-\Omega)/N)}$$
and again, after a time average
$$A_P = {1\over8\sin^2\left(\pi (m'-m-\Omega)\over N\right)}.$$
This result should be compared to $(2)$. For $m,m',\Omega\ll N$, the
polymer rate of absorption approximates that of the Fock oscillator up
to a factor due to replacing an integral over $[0,2\pi]$ by a sum over
$N$ values.

However, as soon as we drop this restriction, there is a large deviation
and, in fact, the rate of absorption is periodic in $\Omega$ with
period $N$. Thus, for fixed $N$ there is a large deviation between the
two expressions at large frequencies. 

Another difference between the two cases is that while all natural
numbers $m$ appear as frequencies in the Fock case, for fixed $N$, only
its divisors appear as ``frequencies'' (inverse periods) of
$|\psi_m\rangle$. Thus the total spectrum for transitions between all
possible $|\psi_m\rangle$ for fixed $N$ depends on number theoretic
properties of $N$. You would see an absorption line in the spectrum
only for frequencies $\Omega$ which can be written as the difference
of two divisors of $N$ (plus multiples of $N$). For example the
spectrum would have only one transition line (up to periodicity
$\Omega$ with period $N$) for $N$ prime compared to an $N$
which has many small divisors.

We should point out that the states $|\psi_m\rangle$ are not the most
general vectors in the polymer Hilbert space. However, one does not
gain ``resemblence'' with the Fock space situation by studying more
general vectors: What is important for the (time averaged) absorption
rate is just the periodicity in time of the support of the state and
in this respect the vectors $|\psi_m\rangle$ cover already all
possibilities.

\chapter{Appendix B: Several oscillators}
Our result above that at least for large $N$ the polymer absorption
rate can look like the Fock rate can be traced back to the fact that
for the harmonic oscillator the time evolution is essentially a $U(1)$
which can be approximated sufficiently by a $\MZ_N$. This however is
not possible for finite subgroups of $\MR$ which becomes relevant for
the time evolution once several oscillators are involved.

Thus now we want to consider two oscillators with not necessarily
identical frequencies. Furthermore we assume there is a linear
coupling $\lambda x_1 x_2$ where $\lambda$ is the coupling constant. 
The combined system can be diagonalised in terms of eigenmodes with
eigenfrequencies depending on $\lambda$. If the ratio of these
eigenfrequencies is irrational the time-evolution is no longer
periodic but only quasi-periodic.

In the polymer description the total Hilbert space is now the tensor
product of two polymer spaces for the two eigenmodes and the time
evolution acts as $U(t)|z_1\rangle\otimes|z_2\rangle=|e^{i\omega_1
t}z_1\rangle\otimes|e^{i\omega_2t}z_2\rangle$. If we couple the
radiation field only to the first oscillator, say, it couples to both
eigenmodes unless $\lambda=0$. As we saw in the treatment of the
single oscillator, the coupling only happens at discrete times $t'$ when
there is an overlap between the time evolved initial and final
state. For two coupled oscillators there has to be an overlap for both
eigenmodes, otherwise the matrix element vanishes. If the two
eigenfrequencies have an irrational ratio, the mutual overlap happens
only a finite number of times (over all $t'$ not just one period) and
thus the absorption eventually has to stop. Thus all absorption lines
of these coupled oscillators in any state will eventually go away in
the polymer description. 

Strictly speaking, one should not only consider states which are
supported at finitely many $|z\rangle$ as one could as well have
countably many. In that case, the coefficients have to be square
summable which also means that even if there could be absorption at
arbitrarily late times the strength will eventually have to go to
zero. 

This conclusion hinges on the assumption of incommensurable
frequencies. Otherwise the time evolution is periodic and thus similar
to the case of a single oscillator. One might
object that the question if $\omega_1/\omega_2$ is in $\MQ$ or in
$\MR\setminus \MQ$ cannot be established by an experiment of finite
precision. Note however that the scalar product in the polymer Hilbert
space exactly requires such infinte precision as otherwise two
prepared states would overlap only with probability zero. Thus to have
interactions at all in the polymer case one has to concede infinite
precision in preparation of the experiment. Furthermore, the
absorption will be discontinuous in $\lambda$ as the ratio of
eigenfrequencies depends on it.

\closebib

\chapter{References}
\bigskip
\parindent =2cm
\eprint{HP}{Helling, Robert C. and Policastro, Giuseppe}{String quantization: Fock vs. LQG representations}{hep-th/0409182}{}{Cited: 1 1 1 1 1 2 2 3 4 5 6 6 }
\eprint{LOST}{Lewandowski, Jerzy and Okolow, Andrzej and Sahlmann, Hanno                  and Thiemann, Thomas}{Uniqueness of diffeomorphism invariant states on holonomy-                  flux  algebras}{gr-qc/0504147}{}{Cited: 4 }
\eprint{NP}{Nicolai, Hermann and Peeters, Kasper}{Loop and spin foam quantum gravity: A brief guide for                  beginners}{hep-th/0601129}{}{Cited: 1 }
\paper{NPZ}{Nicolai, Hermann and Peeters, Kasper and Zamaklar, Marija}{Loop quantum gravity: An outside view}{Class. Quant. Grav.}{22}{2005}{R193}{{\tt hep-th/0501114} }{Cited: 1 }
\paper{T1}{Thiemann, Thomas}{The LQG string: Loop quantum gravity quantization of string                  theory. I:  Flat target space}{Class. Quant. Grav.}{23}{2006}{1923--1970}{{\tt hep-th/0401172} }{Cited: 1 1 2 2 5 }
\eprint{T2}{Thiemann, Thomas}{Loop quantum gravity: An inside view}{hep-th/ 0608210}{}{Cited: 2 }

\bye